\documentclass[aps,jcp,twocolumn]{revtex4-1}
\usepackage{amsmath}
\usepackage{amsfonts}
\usepackage{graphicx}
\usepackage{epstopdf}

\makeatletter
\DeclareFontFamily{OMX}{MnSymbolE}{}
\DeclareSymbolFont{MnLargeSymbols}{OMX}{MnSymbolE}{m}{n}
\SetSymbolFont{MnLargeSymbols}{bold}{OMX}{MnSymbolE}{b}{n}
\DeclareFontShape{OMX}{MnSymbolE}{m}{n}{
    <-6>  MnSymbolE5
   <6-7>  MnSymbolE6
   <7-8>  MnSymbolE7
   <8-9>  MnSymbolE8
   <9-10> MnSymbolE9
  <10-12> MnSymbolE10
  <12->   MnSymbolE12
}{}
\DeclareFontShape{OMX}{MnSymbolE}{b}{n}{
    <-6>  MnSymbolE-Bold5
   <6-7>  MnSymbolE-Bold6
   <7-8>  MnSymbolE-Bold7
   <8-9>  MnSymbolE-Bold8
   <9-10> MnSymbolE-Bold9
  <10-12> MnSymbolE-Bold10
  <12->   MnSymbolE-Bold12
}{}

\let\llangle\@undefined
\let\rrangle\@undefined
\DeclareMathDelimiter{\llangle}{\mathopen}%
                     {MnLargeSymbols}{'164}{MnLargeSymbols}{'164}
\DeclareMathDelimiter{\rrangle}{\mathclose}%
                     {MnLargeSymbols}{'171}{MnLargeSymbols}{'171}
\makeatother

\newcommand{\inner}[2]{\left< #1 \vphantom{#2} \right|
 \left. #2 \vphantom{#1} \right>} 
\newcommand{\braket}[3]{\left< #1 \vphantom{#2#3} \right|
 #2 \left| #3 \vphantom{#1#2} \right>} 
 
\newcommand*{\pd}[2]{\frac{\partial{#1}}{\partial{#2}}}

\newcommand*{\avg}[1]{\left\llangle #1 \right\rrangle}

\newcommand*{\erf}{\operatorname{erf}}

\begin{document}

\title{ Fluctuation Theory of Ionic Solvation Potentials}

\author{ David M. Rogers}
\affiliation{ University of South Florida, 4202 E. Fowler Ave., CHE 205, Tampa, FL 33620}
\email{ davidrogers@usf.edu}

\date{\today}


\begin{abstract}
%
  This work presents a rigorous statistical mechanical theory of solvation free energies,
specifically useful for describing the long-range nature of ions in an electrolyte solution.
The theory avoids common issues with field theories by writing
the excess chemical potential directly as a maximum-entropy
variational problem in the space of solvent 1-particle density functions.
The theory was developed to provide a simple physical picture of
the relationship between the solution's spatial dielectric function,
ion screening, and the chemical potential.
The key idea is to view the direct correlation function of molecular
Ornstein-Zernike theory as a Green's function for both longitudinal
and transverse electrostatic dipole relaxation of the solvent.
Molecular simulation data is used to calculate these direct correlation functions,
and suggests that the most important solvation effects can be captured with only a
screened random phase approximation.
Using that approximation predicts both the Born solvation free energy
and a Debye-H\"{u}ckel law in close agreement with
the mean spherical approximation result.
These limiting cases establish the simplicity and generality
of the theory, and serve as a guide to replacing local dielectric and
Poisson-Boltzmann approximations.
\end{abstract}

\maketitle

\section{ Introduction}

  The importance and many applications of implicit solvent methods hardly requires an introduction.
A complete equilibrium answer to this problem would be given by a computable expression
for the free energy of transferring a set of molecules with a fixed configuration into solution.
However, it is more common (and intuitive) to present the solvation free energy in terms
of a coarse-grained solvent density.  This recognizes both that the solvent has special, `relevant,'
degrees of freedom and that they respond to the solute in a self-consistent way.

  Most heavily used solvation methods focus on electrostatics, carrying
over the Maxwell theory to atomistic volumes by assuming the solvent
dipole density responds to a local electric
field with a slope that is essentially the solvent dielectric coefficient,
$\epsilon_r \sim 80$ in water.\cite{pren12,bmenn97,mschn07,tduig13}
For ionic solutions, the theory is extended with a mean-field assumption for the ion
densities that leads to the Poisson-Boltzmann equation and the Debye-H\"{u}ckel
(DH) limiting law for ionic solvation free energies.  If these theories are used as a starting point,
the goal is to find transferrable models for predicting the spatial behavior of $\epsilon_r(k)$,
and corrections to ion densities, from which all other quantities are derived.
However, comparing theory and experiment within this paradigm requires
contorting direct measurements from both sides to find ``the dielectric,'' which can
depend sensitively on boundary conditions.

  Experimentally, the procedure for measuring dielectric response is well-defined
on a macroscopic scale,\cite{rbuch01} and corresponding molecular dynamics
calculations of spatial dielectrics, $\epsilon_r(k)$,
have appeared for many bulk liquids.\cite{mskaf95,iomel97,pbopp96,pbopp98}
The latter show that the manipulations required to transform
dipole-dipole or current-current correlation functions, $\Sigma(k)$, into $\epsilon_r(k)$
have the form $\Sigma / (\beta^{-1} - \Sigma/\epsilon_0)$ (where $\beta = 1/k_B T$
is the thermal energy scale), and can exacerbate numerical errors
-- even though solvation energies scale as $\beta \Sigma$.
Calculations for geometries including surfaces have also appeared,\cite{hster03,vball05}
as well as methods aimed at computing $\epsilon_r$ in real space as a function of
distance from a solute.\cite{azase01,afrie11,cscha15,cscha16}

  These latter calculations highlight the issues faced by turning the Maxwell theory,
meant for describing macroscopic scales, into an atomistically
detailed calculation device.  First and foremost, the picture of electromagnetic
wave reflection at a macroscopic dielectric interface does not scale down to an
atomistic theory.\cite{dmaty14}  It is well-known that solvent dipoles are
oscillatory at an interface due to orientational saturation,\cite{vball05} and that water has
heightened in-plane correlations near hydrophobic solutes.\cite{afrie11}
These boundary layer effects motivate treatment of interfacial water
as a chemically distinct species.\cite{bbagc01,dmart14,dmart17}
Describing the electric field experienced by molecular solutes can be
done within the Maxwell theory only by altering to the usual macroscopic
boundary conditions to account for these effects.\cite{mheyd12}

  This paper explores a theory of solvent density response that is distinct
from the Maxwell dielectric picture.  It eliminates the solvent dielectric in
favor of a direct prediction of the local force (including electric and molecular fields)
to which solvent and solute dipoles respond.  The approach mirrors Ref.~\citenum{rrems16}
and the results are closely tied to integral theories of solution.\cite{dbegl97}
We deviate from those theories by only
describing solvent response, with a focus on electrostatics.
The major results are summarized in section~\ref{s:latt} with a sketch
presenting solvation free energy components for an exactly solvable model.
In sec.~\ref{s:gen}, we present the general theory by
expressing the free energy in terms of a density functional,
explaining our method for calculating correlation functions of real fluids (\ref{s:corr})
and then demonstrating their use in determining the potential distribution (\ref{s:pot}).
Section~\ref{s:res} shows two results.  First, the effective local field (molecular potential
or direct correlation function) is computed for electrolyte solutions of varying concentrations.
These show that the contribution of nearby charges is screened out almost
exactly following an error-function like splitting.
Second, sec.~\ref{s:limit} shows that, with the error-function screening, Born and Debye-H\"{u}ckle
solvation laws are recovered even using a random phase-like approximation.
The discussion in sec.~\ref{s:disc} compares these results with recent literature,
and we conclude by listing some of the unexplored consequences of the present development.

  Our final results support the molecular Ornstein-Zernike 
perspective that solvent position and orientational distributions should be described by
effective energy functions in close connection to pair correlations.
This is the same conclusion as an earlier nonlocal response function theory
developed for describing electron transfer.\cite{nrft,dmart08}
The present work was developed independently,
and is more simply motivated by the structure of the inverse pair correlations.
Our main result constitutes a novel and rigorous foundation for using density functional
methods to replace dielectric solvation theories for calculating free energies.
The focus on the dielectric is appropriate because, even using the mean spherical approximation
(MSA) solution for the primitive model of electrolytes,
the state-dependence of the solvent dielectric constant
is a major difficulty when comparing to experiment.\cite{rperr88,akova00,bmari12}
Recent progress in integral theories has been made in predicting nontrivial
density-dependence for solvent dielectric response.\cite{frain01,kdyer08}
Our approach of analyzing correlation functions has close connections
to both dipolar fluctuation theories\cite{mskaf95,iomel97,pbopp98,vball05,cscha15,cscha16}
and inference on bridge functions for RISM models.\cite{sover12,szhao13,gchue14,sshen16}
However, it differs from both in its focus on predicting solvent
density distributions directly through a {\em simple} maximum entropy procedure, as
opposed to parameterizing any particular theory.

\section{ Lattice Model}\label{s:latt}

  We will sketch our main results by recalling the simple, exactly solvable
system made up of a lattice of polarizable point dipoles, $p_i$,
at fixed positions, $r_i$, in the presence of fixed ions.\cite{snovi95,sravi96,apapa97,hster03}
Our purpose is to ground the discussion by demonstrating that this system contains
all electrostatic contributions to the solvation free energy --
dielectric self-energy, screening and solvent dispersion energies.
The result also reveals the root cause of some issues with ``local dielectric'' models.
Formulas which apply only to this system will have the superscript ``ref.''
As a matter of convenience, statistical mechanical averages are written
using double-angle brackets, as in $\avg{p}$ for the average of the dipole vector, $p$.
Single angle brackets are reserved for a bra-ket notation for matrix inner products
(defined in the Appendix~\ref{s:ip}).

  The long-range part of the electrostatic energy of the system can be written as,\cite{uessm95,atouk00}
\begin{equation}
\begin{split}
E_\text{int}^\text{ref} = \frac{1}{2} \sum_{ij} (q_i + p_i \cdot \nabla_i) (q_j - p_j\cdot\nabla_i) G(r_i - r_j) \\
- \frac{\eta}{4\pi\epsilon_0 \sqrt{\pi} } \left( \frac{2\eta^2}{3} \sum_i |p_i^2| + \sum_i q_i^2 \right)
\end{split} \label{e:ref}
.
\end{equation}
In order to include the self ($i=j$) term, we only consider the long-range part of the
Coulomb potential, so that
\begin{equation}
G(r) = \sum_n \frac{\erf(\eta |r + n|)}{4\pi\epsilon_0 |r + n|}
,\label{e:G}
\end{equation}
where the sum runs over 3D lattice vectors, $n$, of the system's unit cell.
For a uniform distribution of dipoles, the screening effectively
ignores solvent near each molecule -- creating a molecular cavity around each one.
If we assume the dipoles can take any magnitude with internal energy given by,
\begin{equation}
E^\text{ref}_0 = \frac{1}{2} \sum_i |p_i|^2 / \alpha_i
,\label{e:ref0}
\end{equation}
then the potential energy function for the collection of
dipole moments, $p$, is equivalent to that of a forced Harmonic oscillator.

  Since this system is Gaussian, the dipole fluctuations ($\Delta p_i \equiv p_i - \avg{p_i}$)
exactly satisfy the matrix equality,
\begin{equation}
[\Sigma^{-1}]_{ij}/\beta = [G_p]_{ij} + \delta_{ij}/\alpha_i
  ,\label{e:diel}
\end{equation}
for any position of the ions.
Here $\Sigma_{ij} = \avg{ \Delta p_i \Delta p_j^T }$ is the
dipole-dipole correlation function, $\rho_w$ is the density of dipoles, and
$[G_{p}]_{i,j} \equiv -(1-\delta_{ij})\nabla_i \nabla_i^T G(|r_i-r_j|)$ is
the interaction energy between dipoles at $r_i$ and $r_j$
(The Kronecker delta function omits the $i=j$ terms).
Eq.~\ref{e:diel} gives a trivial derivation of the formula for the spatial dielectric response of a medium.
Typically, one would infer the dielectric function, $\epsilon \equiv \epsilon_0 \epsilon_r$,
by comparing the linear response predicted by
statistical mechanics, $\Delta p = \beta \Sigma {\vec E}^\text{ext}$,
to the phenomenological equation expressing polarization
by a local electric field, $\rho_w \Delta p = \epsilon (\vec E^\text{ext} - G_p\Delta p)$.
Then the comparison reads,
\begin{equation}
\rho_w \Delta p = \epsilon (\vec E^\text{ext} - G_p \Delta p) = \beta \rho_w \Sigma \vec E^\text{ext}
\label{e:compar}
\end{equation}
and is solved when $\rho_w \epsilon^{-1} = (\beta \Sigma)^{-1} - G_p$.
Substituting the exact correlation function from Eq.~\ref{e:diel} shows that the phenomenological equation
gives the diagonal matrix, $\epsilon_{ij}^{-1} = \delta_{ij}/\rho_w \alpha_i$.
If, in addition to electrostatic interactions, there are other local interactions between dipoles,
then $(\beta \Sigma)^{-1} - G_p$ will still be localized.  However, its inverse will not.
This non-locality is the central problem with adapting the Maxwell theory of dielectric response
to atomistic scales.

  This simple, extensively studied,\cite{hzhou92,sravi96,apapa97}
picture of a lattice of dipoles provides a new perspective on the role of dielectric
in integral theories of electrolyte solutions.
The direct correlation functions between water dipoles
provide the effective interaction energy between dipoles at every separation in a fluid,
and are more transferrable than their inverse ``dielectric.''
The solvent response part of dielectric theory is replaced with a rigorous
maximum entropy theory for predicting the solvation free energy.
The density functional to be maximized is the natural target for
developing approximate physical theories.
If the interactions are restricted to long-range,
avoiding large high-wavenumber perturbations,
then a Gaussian approximation is shown to be very accurate.

  To finish the simple example, note that the classical free energy for any configuration of ions,
$A^\text{ref}(\beta, r_\alpha)$, can be found by integrating the partition function
over the vector of solvent dipoles,
\begin{align}
A^\text{ref}(\beta, r_\alpha) &= \tfrac{1}{2\beta} \ln |\Sigma^{-1} / 2\pi| + E_\text{eff}^\text{ref}
\label{e:FEref} \\
\intertext{In the limit of a uniform distribution of $N$ dipole locations and letting
$\eta \to\infty$ while $1/\bar\alpha = 1/\alpha - \eta^3/3\epsilon_0\pi^{3/2}$,}
\ln |\Sigma^{-1}/2\pi| &= 3N \ln (\beta / 2\pi \bar\alpha) \notag \\
  &\quad + (N-1) \ln (\epsilon_r)
  + \ln \left|I_3 + \bar\alpha \rho_w J \right| \label{e:disp} \\
E_\text{eff}^\text{ref} &= \frac{1}{2\epsilon_r} \sum_{ij} q_i q_j G^\text{ref}(r_i - r_j) 
     - \frac{\eta}{4\pi\epsilon_0\sqrt{\pi}}\sum_i q_i^2 \label{e:Eeff} \\
\epsilon_r &= 1 + \bar\alpha \rho_w/\epsilon_0 \label{e:epsr}
.
\end{align}
Here $J$ is the depolarization tensor that expresses the surface energy,
$E_\text{surf} = M^T J M/2V$, due to a net system dipole, $M$,
surrounded by a bulk medium of dielectric $\epsilon_e$.\cite{vball14}
For a spherical boundary at infinity, $J = I_3/\epsilon_0 (2\epsilon_e + 1)$,
with $I_3$ the $3\times 3$ identity matrix.

  Eq.~\ref{e:FEref} contains electrostatic screening, dielectric self-energy and solvent
dispersion energies.
The screening appears directly in $E_\text{eff}$ via $\epsilon_r > 1$.
The dielectric self-energy is the free energy of solvation for a single ion.
It appears as the difference between the $i=j$ term,
where $G_\text{ref}$ is $\eta/2\epsilon_0 \pi^{3/2}$,
and the right-hand side.
Identifying this with the Born solvation free energy lets us make the definition
$R_\text{B} \equiv \sqrt{\pi}/2 \eta$.
The solvent dispersion energy appears in the
normalization constant, $|\Sigma^{-1}|$.\cite{sjohn11}
We have left the mass contribution out of ~\ref{e:disp}
and present only the classical free energy.
Adding ions does not remove polarizable centers in this picture, so there is no dispersion energy
of solvation unless the ion polarizability differs from bulk water.

  The result is exact for the reference system, but cannot be generalized to solution
density functions because integration over all possible densities is ill-defined.
Physically, the space of all possible density functions is much larger
than the configuration space of the system and introduces non-physical degrees of freedom.\cite{fdyso72}
The Hubbard-Stratonvich transformation provides one alternative route,
but also meets difficulties with integration over infinite field variables.
Even when those integrations can be defined with a convergent limit,
the field functional cannot be given in a closed form and must be approximated
in a way directly paralleling density functional theories.\cite{adieh97}
It should be noted that this procedure has been carried out rather clearly
for fluids with soft pairwise interactions in Ref.~\citenum{jmart16},
and both it and an earlier work\cite{zwang10}
contain parallels to many of the results from our Gaussian approximation
in Sec.~\ref{s:limit}.  Implementations of that theory have
also helpfully cast the solution process as a maximum entropy problem.\cite{jpujo15}

  This work avoids issues with field variable integration by applying large deviation theory to
find a representative density that yields the exact distribution of solute interaction energies.
This approach appears to be a new alternative to the random phase approximation route
to electrolytes solution structure recently shown by Frydel and co-workers.\cite{dfryd16,yxian17}
Comparing the results of this procedure with the exact Eq.~\ref{e:FEref} is interesting
because it helps eliminate misconceptions and provides an intuitive
context for all of the quantities that will appear below.

\section{ General Model}\label{s:gen}

\begin{figure}
{\centering 
\includegraphics[width=0.49\textwidth]{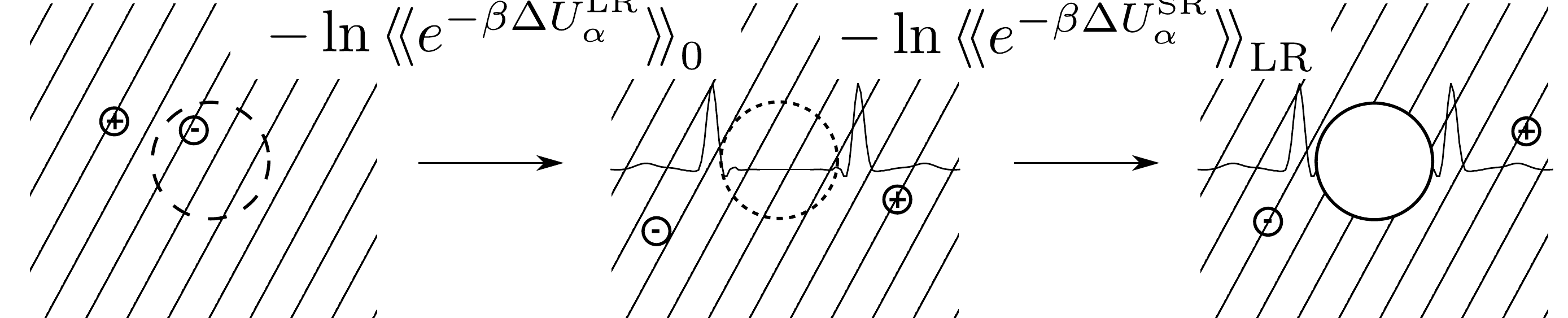}
\caption{Division of the excess chemical potential into long-range polarization (left) and short-range chemical (right) steps.  On the left, the solute and solvent are not interacting.  In the center,
the solute-solvent interaction energy is given by Eq.~\ref{e:uLR} and at the right the solvent and
solute are fully interacting.
This cycle was featured in Ref.~\citenum{rrems16}.}\label{f:cyc}
}
\end{figure}

  The molecular picture of a uniform dipolar fluid above works well because
local interactions were screened out by the replacement $1/r \to \erf(\eta r)/r$.
Locally, the system behaves like a fluid with only short-range order while we focus on treating
long-range correlations correctly.  This leads us to the thermodynamic cycle of Fig.~\ref{f:cyc}.
Solvation of a charged species is divided into a long-range (LR) and a local (SR)
contribution.  The long-range step is performed {\em first}.
According to the potential distribution theorem, the excess chemical potential of
a solvated molecule, $\alpha$, is then expressed exactly as,
\begin{align}
\beta \mu^\text{ex}_\alpha &= -\ln \avg{ e^{-\beta \Delta \hat U^\text{LR}_\alpha} }_0 - \ln \avg{ e^{-\beta \Delta \hat U^\text{SR}_\alpha} }_\text{LR} \label{e:muex} \\
\Delta \hat U^\text{LR}_\alpha &\equiv \braket{ \hat\rho_q(r') }{ \frac{\erf(\eta_\alpha |r' - r|)}{4\pi\epsilon_0 |r' - r|}}{\hat \rho_\alpha(r) } \label{e:uLR} \\
\Delta \hat U^\text{SR}_\alpha &\equiv \Delta \hat U_\alpha - \Delta \hat U^\text{LR}_\alpha .
\end{align}
The first term on the right of Eq.~\ref{e:muex} is defined as $\beta\mu^\text{ex,LR}_\alpha$,
the focus of this work.
Here, $\hat\rho_q(r)$ is the instantaneous charge density at point $r$
corresponding to a single solvent
microstate, while $\rho_q(r)$ is the average charge density which is zero by symmetry.
Eq.~\ref{e:uLR} for the long-range interaction energy uses a bra-ket notation
for the double-integral over $dr$ and $dr'$.
Hats are used used on quantities that depend on the molecular coordinates.
Although this long-ranged interaction just contains the screened Coulomb energy,
we note that it is simple to generalize $\eta_\alpha$ to $\eta_\alpha(r)$, so that
different atoms can have different effective radii.  The total solvation free energy,
$\mu^\text{ex}_\alpha$, will not be affected by the choice of $\Delta \hat U_\alpha^\text{LR}$,
but our results confirm that choosing $\sqrt{\pi}/2\eta = R_B$
minimizes the importance of the second, short-range step.
The theory in Sec.~\ref{s:pot} is general enough to handle any
choice for $\Delta \hat U^\text{LR}_\alpha$, including the full
$\Delta \hat U_\alpha$ itself.

\subsection{ Correlation Functions}\label{s:corr}

  We have recently presented a simple Fourier-space method
for inferring the direct correlation function from molecular dynamics data.\cite{droge17e}
This method of inverting the dipole correlation function to find an ``effective
dipole-dipole interaction energy'' is synonymous with computing the direct correlation
function of  Ornstein-Zernike theory and the polarization structure factor of nonlocal
response function theory.\cite{nrft,dmart08}
Formally, define a vector, $F_\alpha$, characterizing the orientation of a molecule of type $\alpha$.
For example, the vector characterizing
a molecule containing a point dipole could be $F = [1, p_x, p_y, p_z]^T$.
Correlations between the density operators,
\begin{equation}
\hat F_\alpha(r) = \sum_{j=1}^{N_\alpha} F_{\alpha,j} \delta(r - r_j)
,
\end{equation}
where $j$ indexes the $N_\alpha$ molecules of type $\alpha$,
then report on both scalar and vector interactions.

\begin{figure}
{ \centering
\includegraphics[width=0.45\textwidth]{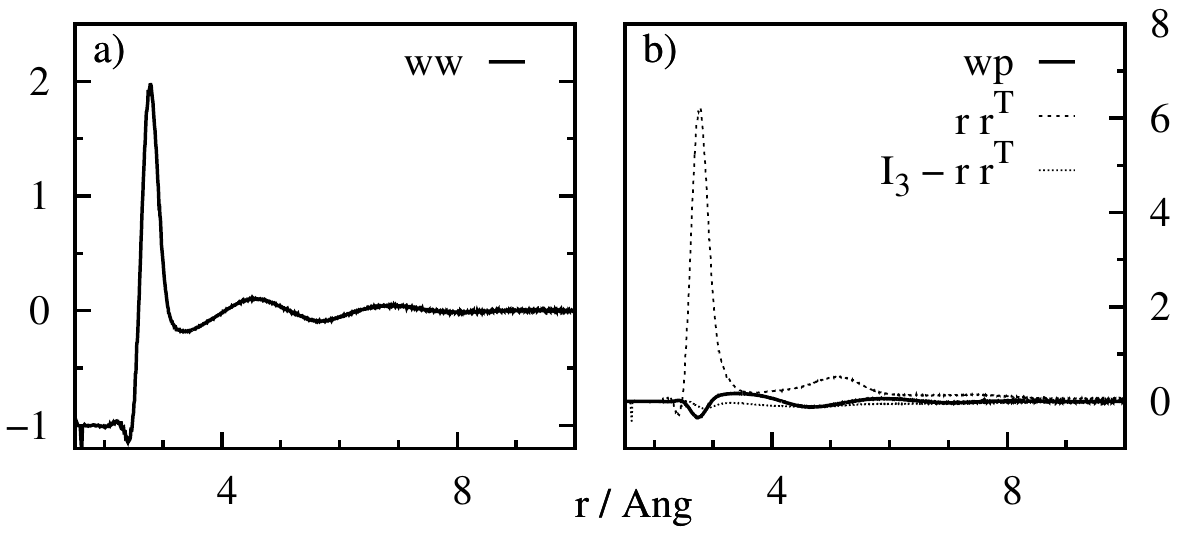} \\
\includegraphics[width=0.45\textwidth]{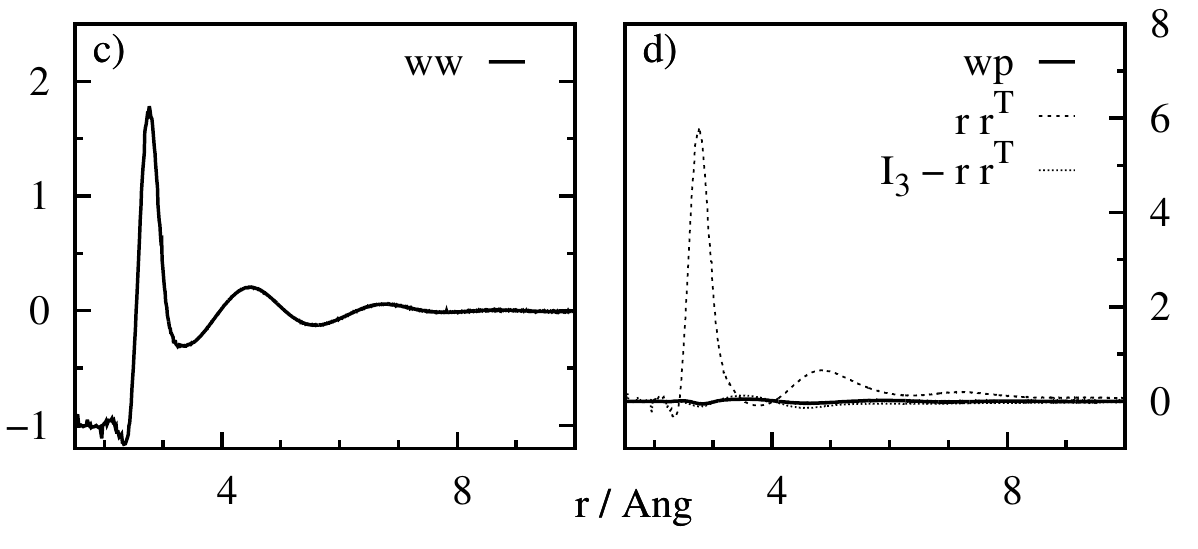}
\caption{Radial correlation functions ($h(r) = g(r)-1$) for water oxygen and dipole correlation functions.
The top row (a-b) shows results from the SPC/E water model.
The bottom row (c-d) shows results of the TIP5P water model.
Water-dipole correlation functions (labeled wp) have units of $e_0$\AA{}.
The remaining lines in b and d are dipole-dipole correlations with radial ($\hat r \hat r^T$) or tangential
symmetry ($I_3 - \hat r \hat r^T$) and have units $(e_0\text{\AA{}})^2$.}\label{f:water}  }
\end{figure}

  The correlation function for these vectors is organized into a matrix where
molecule indices are combined with vector indices and Fourier transformed,
\begin{equation}
Q_{\alpha\gamma}(k) \equiv \int_V e^{-i r\cdot k} \avg{ \Delta \hat F_\alpha(r) \Delta \hat F_\gamma(0)^T } dr
.\label{e:Q}
\end{equation}
The integral ranges over the unit cell with volume $V$.
To minimize error in the low-frequency components and to compute $Q$ efficiently,
Eq.~\ref{e:Q} is estimated from simulation data by
averaging squared Fourier transforms ($\hat \psi_\alpha(k) \equiv \mathcal F[\hat F_\alpha(r)](k)$),
so that $Q = \avg{\Delta \hat \psi \Delta \hat\psi^\dagger}/V$.
Also, we define a potential after dividing by the volume
so $\phi_\alpha(k) \equiv \mathcal F[ \Phi_\alpha(r)](k) / V$.
The Appendix gives helpful relations for the Fourier transform used here.

  The correlation function is related to the Ornstein-Zernike definition,\cite{pdt6}
\begin{equation}
Q_{\alpha\gamma}(k)^{-1} = \pd{\beta\mu^\text{id}_\alpha}{\rho_\gamma} - c_{\alpha\gamma}(k)
.\label{e:iQ}
\end{equation}
Here,
\begin{equation}
\beta \mu^\text{id}_\alpha = \ln (\rho_\alpha \Lambda^3_\alpha / q^\text{int}_\alpha)
\end{equation}
is the ideal gas expression for the chemical potential of species $\alpha$.\cite{pdt2}
Approximations to $Q$ are available both from analysis of experimental data\cite{pbopp98},
and from solutions to integral equations for charged or dipolar hard-sphere liquids.\cite{mwert71,mwert84,nrft} 
The results section presents the correlation functions and inverses computed
from MD simulations of a series of 1:1 electrolyte solutions.

\subsection{ Solvation Potential Distribution}\label{s:pot}

  The basic quantity of the present theory is an exponential average,
\begin{equation}
Z[\Phi ; V,\beta] = \avg{\exp{\inner{\hat \rho}{\Phi}}}
,\label{e:Z}
\end{equation}
which is taken over the grand-canonical ensemble with $\mu$
describing the system before coupling to the solute (when $\Phi = 0$).
A constant shift of $\Phi$ has the same effect as changing the
chemical potential in Eq.~\ref{e:Z}.
The inner product in the exponent is defined in Appendix~\ref{s:ip} and carries units of volume
if both sides are in real space and inverse volume if both sides are in Fourier space.
The long-range excess chemical potential can be
written in terms of Eq.~\ref{e:Z} as
$\beta\mu^\text{ex,LR}_\alpha = -\ln Z[-\beta \Phi_\alpha ; n,V,\beta]$,
which requires Legendre transformation of $\ln Z$ into the nVT ensemble,
and defines $\Phi_\alpha$ to be the pair interaction field produced by the solute
(by comparison to Eq.~\ref{e:uLR}) as,
\begin{equation}
\Delta \hat U^\text{LR}_\alpha = \inner{ \Phi_\alpha }{ \hat\rho } = \inner{ \phi_\alpha }{\hat\psi}
.\label{e:pLR}
\end{equation}

\begin{figure}
{ \centering
\includegraphics[width=0.45\textwidth]{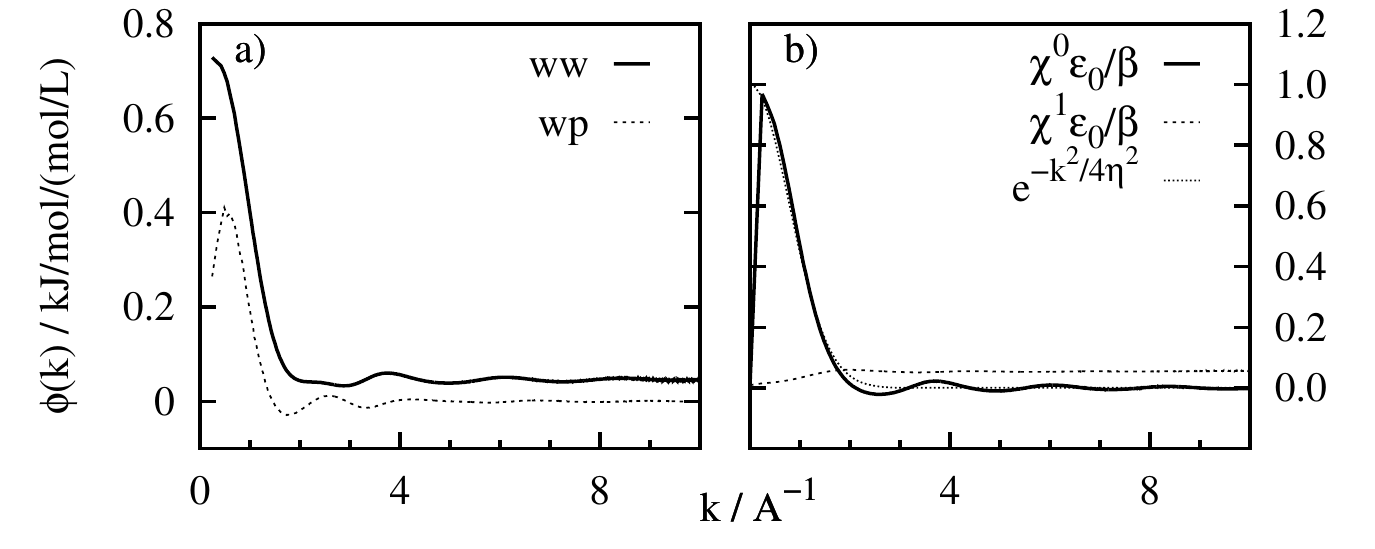} \\
\includegraphics[width=0.45\textwidth]{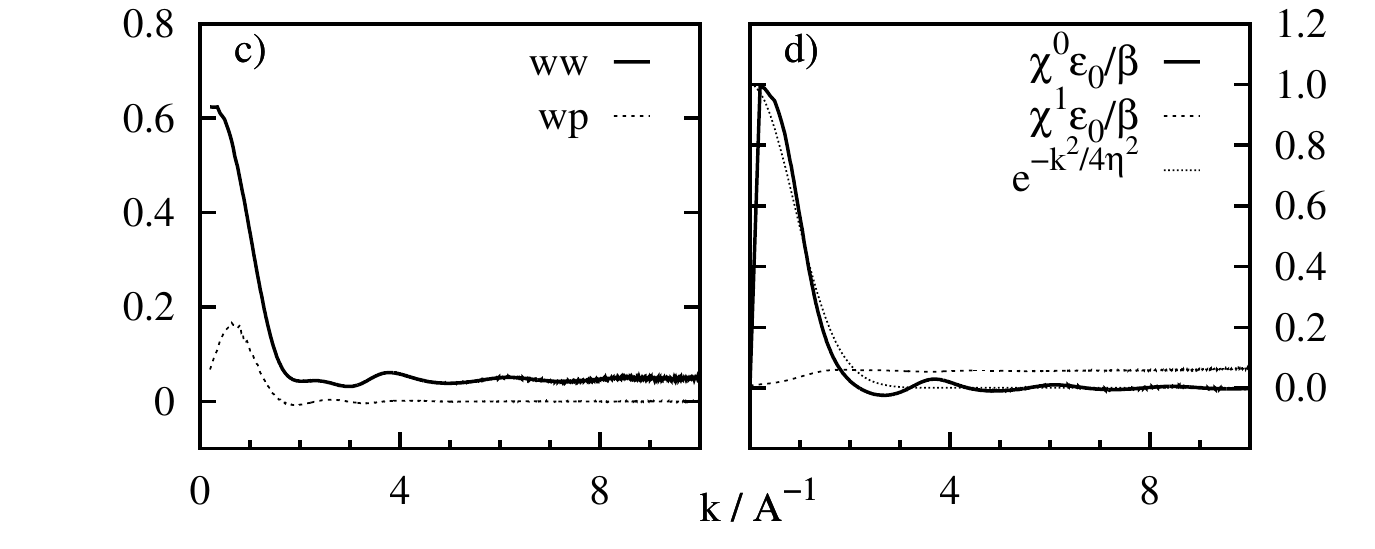}
\caption{Inverse correlation functions in reciprocal space
for water center of mass (left) and dipole correlation (right) functions.
Note the units of the water-dipole correlation function contain an extra $1/e_0$\AA{}.
The top row (panels a,b) shows results from SPC/E and the bottom
row (panels c,d) shows TIP5P results.  See Eq.~\ref{e:chi}
for definitions.  The value of $\eta$ plotted in panel b (SPC/E) is
$1/1.8$\AA{}  and in panel d (TIP5P) is $1/1.6$\AA{}.}\label{f:iwater}  }
\end{figure}

  Equation~\ref{e:Z} is a moment generating function for the density field, $\hat \rho$,
and we can define its corresponding density functional as the Legendre transform,
\begin{equation}
\mathcal R[\rho ; V,\beta] \equiv \inf_{\Phi} \left[ \ln Z[\Phi ; V,\beta] - \inner{\rho}{\Phi} \right]
.\label{e:R}
\end{equation}
Eq.~\ref{e:R} is a generalized entropy, and is also called a (negative) rate function
for the empirical distribution, $\avg {\hat\rho}_{V,\beta,\Lambda}$
in large deviation theory.\cite{svara08,htouc09}
It is also the negative of the traditional ``density functional.''\cite{dchan86}
The functional $\mathcal R$ is maximized under extra constraints $\Lambda$
at constant volume and temperature when $\rho = \avg {\hat\rho}_{V,\beta,\Lambda}$.
Stated in terms of Fourier-transformed densities and potentials,
$\mathcal R$ satisfies
\begin{align}
\pd{\mathcal R[\psi]}{\psi^\dagger_\alpha(k)} &= -\phi_\alpha(k;\Lambda), \notag \\
V \pd{\phi_\alpha(k; \Lambda)}{\psi_\gamma(k') } &= \delta_{k,k'} Q^{-1}_{\alpha\gamma}(k;\Lambda)
. \label{e:dR}
\end{align}
The second equality requires translational invariance of the constrained ensemble ($V,T,\Lambda$)
and explicitly acknowledges that the fluctuations in Eq.~\ref{e:Q} depend on that ensemble.

  Because $Z$ and $\mathcal R$ form a Legendre transform pair, $\beta\mu^\text{ex,LR}$
is simple to state in terms of $\mathcal R$,
\begin{equation}
-\beta\mu^\text{ex,LR}_\alpha = \sup_\psi \left[ \mathcal R[\psi] - \inner{\psi}{\beta\phi_\alpha} \right]
.\label{e:muLR}
\end{equation}
This equation may prove to be the most useful result in this paper.
It shows that an approximation for Eq.~\ref{e:Z} can be used to create a density
functional that avoids the ``charging''
integration process of density-functional theory.\cite{ssing85,rrems16}
The maximization over densities in Eq.~\ref{e:muLR} can be carried out under
fixed $n = \psi(0)$, but might also include other conditions.
Those other constraints can be used to change the ensemble or
even to fix the density to zero near the origin (as for the mean-spherical approximation).
Because $\mathcal R$ was defined as a minimization problem,
constraints on $\rho$ will be reflected physically by deviations of
the pair potential, $\Phi$, which solves $\mathcal R$ away from the
default ($\rho$-unconstrained) solution at $\Phi_\alpha$.
Rigorously, constrained maximization is justified by the
Gibbs conditioning principle, which states more formal conditions on
the constraint and density spaces.\cite{cleon02}

  This development makes an important departure from traditional Ornstein-Zernike theory.
Rather than seeking to provide an extra closure between direct and indirect correlation
functions, our primary target is to model the rate function (Eq.~\ref{e:R}).
The Ornstein-Zernike relation is embedded in the structure of Eq.~\ref{e:dR},
so that the density fluctuations (radial distribution functions) come out as a consequence
of a proper model for $\mathcal R$.


  So far, the development from Eq.~\ref{e:muex} through Eq.~\ref{e:muLR} has been formal and exact.
To present analytical results in Sec.~\ref{s:limit}, we use the two-moment approximation,
\begin{equation}
\mathcal R[\psi] \simeq - \frac{1}{2} \braket{\Delta \psi}{Q^{-1}}{\Delta \psi}
,\label{e:R2}
\end{equation}
where $\Delta \psi \equiv \psi - \psi_0$ and $\psi_0$
is the (Fourier) density of the uncoupled system.
Technically, the domain of $\psi$ is limited to positive densities,
which could also be included either as a better approximation to $\mathcal R[\psi]$
or as additional constraints.
The results in this work do not include such a constraint.

  Earlier work showed the utility of a Gaussian perturbation theory for the long-range part
of the solvation free energy {\em after} forming a sufficiently large
cavity.\cite{droge08,droge10,afrie11,rrems16}
\begin{equation}
\mu^\text{ex,LR}_\alpha = \avg{ \Delta \hat U^\text{LR}_\alpha }_0
  -  \beta \sigma_\text{LR}^2 / 2 + O(\beta^2)
 .\label{e:mu2}
\end{equation}
Before forming a cavity, the first term in the expansion is zero by symmetry.
We recover this result by inserting Eq.~\ref{e:R2} into Eq.~\ref{e:muLR}, adding $\phi(0)^T \psi(0)$
at $k=0$ to constrain the number of solvent molecules (with undetermined
Lagrange multiplier $\phi(0)$), and maximizing.
This leads to the identifications,
\begin{align}
\Delta \psi(k) &= -\beta Q(k) \phi_\alpha(k) + \delta_{k,0} Q(0) \phi(0)
, \label{e:dpsi} \\
\intertext{and}
-2 \mu^\text{ex,LR}_\alpha &= \beta \sigma_\text{LR}^2
  = \inner{\delta_{k,0} \tfrac{\phi(0) }{\beta}  - \phi_\alpha }{\Delta\psi} \notag \\
   &= \inner{- \phi_\alpha }{\Delta\psi}
       \label{e:s2}
\end{align}
and shows that when Eq.~\ref{e:R2} applies, the linear response
approximation actually gives the ensemble average of the 1-particle density functions
in the coupled state, and proves that the distribution of interaction energies,
$\Delta \hat U^\text{LR}_\alpha$, is also Gaussian.
The last simplification step in Eq.~\ref{e:s2} happens because
each component of the multiplier, $\phi_j(0)$, is chosen so that the corresponding
molecule number does not change ($\Delta \psi_j(0) = \Delta n_j = 0$).

  For describing ionic solvation, specialize the solution components
to a 1:1 electrolyte (e.g. NaCl) in water (modeled as a dipole so $F_\text{H$_2$O}$ is a 3-component
vector).  Each density vector, $\psi(k)$, then has 5 components -- cation and anion
densities, $\psi_q$, plus water dipole vectors, $\psi_p$.
In this basis, the screened interaction energy of an additional ion of charge $z_\alpha$
with solution (Eq.~\ref{e:uLR} and Eq.~\ref{e:pLR})
has the form $\beta \phi_\alpha(k) = z_\alpha f(k) [1,-1,-ik]^T$
where $f(k) \equiv \beta e^{-k^2 / 4\eta^2} / k^2 \epsilon_0$ is the Fourier-transform
of the screened Coulomb operator (Eq.~\ref{e:G}).

\begin{figure}
{\centering
\includegraphics[width=0.45\textwidth]{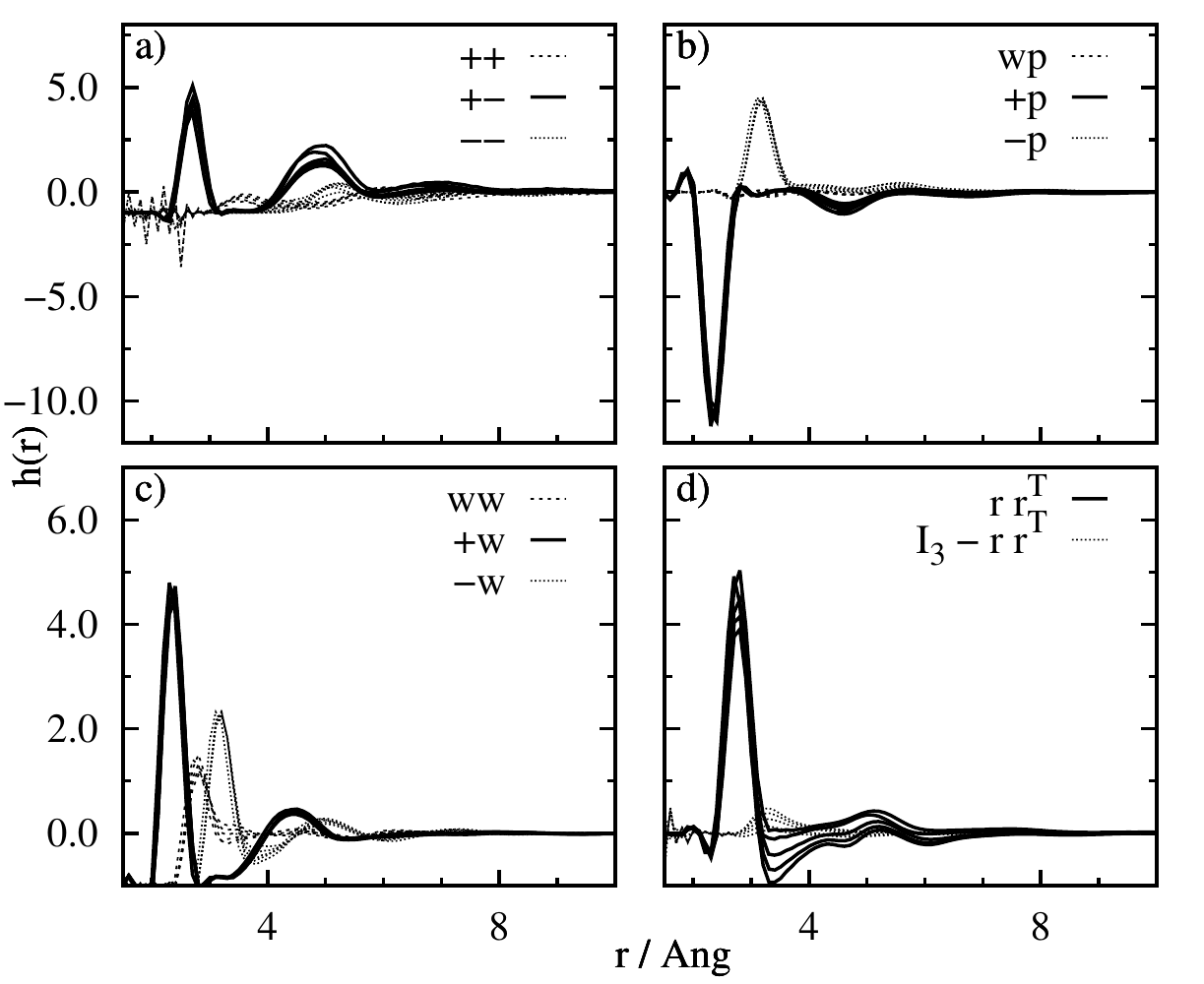}
\caption{Electrolyte radial and dipole correlation functions overlayed
for all five NaCl concentrations.  Labels are as in Fig.~\ref{f:water}.
The right and left scales are identical even though the units differ.
Peaks generally decrease toward zero with increasing concentration
as can be inferred from comparison with corresponding plots in
Ref.~\citenum{sweer03}.  The exception to this trend is the first
minimum of the $\hat r\hat r^T$ component (d), which becomes
more pronounced from 0.5 to 4M.}\label{f:ions} }
\end{figure}

  Equation~\ref{e:dpsi} is the linear response of both water dipoles and ion densities
to charged solutes.  It is given, in the Gaussian approximation,
by the convolution of the indirect correlation functions, $Q$,
with the potential, $\phi_\alpha$.  Since the potential is smooth and long-ranged,
the small $k$ behavior of the correlation functions
is the most important aspect of solvation.  Second most is improving $\mathcal R$
beyond the Gaussian approximation to describe excluded
volume effects.

\section{ Results}\label{s:res}

  We show two key results.  First, for sodium chloride solutions, the direct correlation
functions exhibit error-function like screened forms with little dependence on ionic strength.
Second, the long-range limits of these screened Coulomb forms motivate a random-phase
like approximation whose analytical predictions for solution dielectric and ionic response re-derive
the Born law and extend the Debye-H\"{u}ckel theory with an ionic size-dependence that
closely mirrors the full mean-spherical approximation result.
Together, these results justify abandoning the Maxwell theory and its associated extensions of the
Poisson equation.
Instead, direct charge correlation functions, available from both experiment and simulations,
form the basis for computable methods of solution response and solvation free energies.

\subsection{ Simulation Correlation Functions}\label{s:sim}

  We simulated sodium chloride solutions in SPC/E using the Kirkwood-Buff forcefield model
for ions\cite{mgee11} replicating the simulation
conditions in Ref.~\cite{sweer03} but extending all simulations to at least 10 ns.
Figures~\ref{f:water} and \ref{f:iwater}
show the water position and dipole indirect correlation functions ($F = [1,p_x,p_y,p_z]$)
for pure SPC/E\cite{hbere87} and TIP5P\cite{mmaho00} water models under NVT conditions at 300 K.
These were computed by inverse Fourier-transform of the correlations in reciprocal space.
Deviations of $h(r)$ below the exact value of $-1$ near the origin indicate the amount of numerical error
due to the band-limiting inherent in our method.

  For all vector-scalar interactions plotted in real-space, the geometry should be pictured as
with the vector fixed at the origin, pointing along the $+z$ axis.
Positive interaction energies or low densities then describe unfavorable interactions
of the dipole with molecules on the $+z$ hemisphere of radius $r$.
Vector-vector interactions (e.g. between dipoles $p_1$ and $p_2$)
along a separation direction $\hat r$
in real space are broken into parallel
and isotropic components scaling with
$p_1^T \hat r \hat r^T p_2$ or $p_1^T p_2$, respectively.

  Figure~\ref{f:iwater} plots the components of water's the inverse correlation function, $Q^{-1}(k)$.
Panels b and d should be compared with Figs.~7 and ~15 of Ref.~\citenum{nrft},
after noting that each point, $Q^{-1}(k)$, plotted here is the result of a $4\times 4$
matrix inverse.
The two symmetry components of the dipole-dipole direct
correlation function, $\chi \equiv Q^{-1}_{pp}$,
in Fig.~\ref{f:iwater}b and~\ref{f:iwater}d are,
\begin{equation}
\chi(k) = \chi^0(|k|) \hat k \hat k^T + \chi^1(|k|) (I_3 - \hat k \hat k^T)
.\label{e:chi}
\end{equation}
Since the field produced by a dipole at the origin is proportional to $\hat k \hat k^T$,
$\chi^0$ represents the effective energy governing the
`longitudinal' fluctuations of the second dipole in the direction
parallel to the field lines created by the first water dipole, while $\chi^1$ represents
`free' or `tranverse' fluctuations in the perpendicular direction.
Without screening, these would be $\chi^0\epsilon_0/\beta = 1 + \epsilon_0 / \rho_w \alpha$
and $\chi^1 \epsilon_0/\beta = \epsilon_0 / \rho_w \alpha$ in the lattice model.
With screening, the $1$ in $\chi^0$ is
replaced by the Fourier-transform of the screening charge distribution
while $\chi^1$ is unchanged.
%
This is the usual justification for using $\chi^0(0) = \chi^1(0) = V/\avg{M_x^2}$ (when $J = 0$)
to compute $\epsilon_r(0) = \beta / \chi(0) \epsilon_0 + 1$.
However, away from $k=0$, only $\chi^1$ varies continuously.
Fig.~\ref{f:iwater}b and~\ref{f:iwater}d clearly show that $\chi^0$
starts at 1, but decreases exponentially as $k$ increases.
The fitted line uses $1/\eta = 1.8$\AA{} for SPC/E, which is equivalent to $R_B = 1.6$\AA{}.
Interestingly, the $\chi^1$ component starts near 1/79, but increases to a larger constant
value around $0.0540$ (SPC/E) or $0.0556$ (TIP5P).  Using this in Eq.~\ref{e:epsr}
gives only $19.5$ or $19.0$ for the dielectric, respectively.
This behavior also appears in the MSA solution, and is more pronounced
for dipolar-only solvents.\cite{nrft}

\begin{figure}
{ \centering
\includegraphics[width=0.45\textwidth]{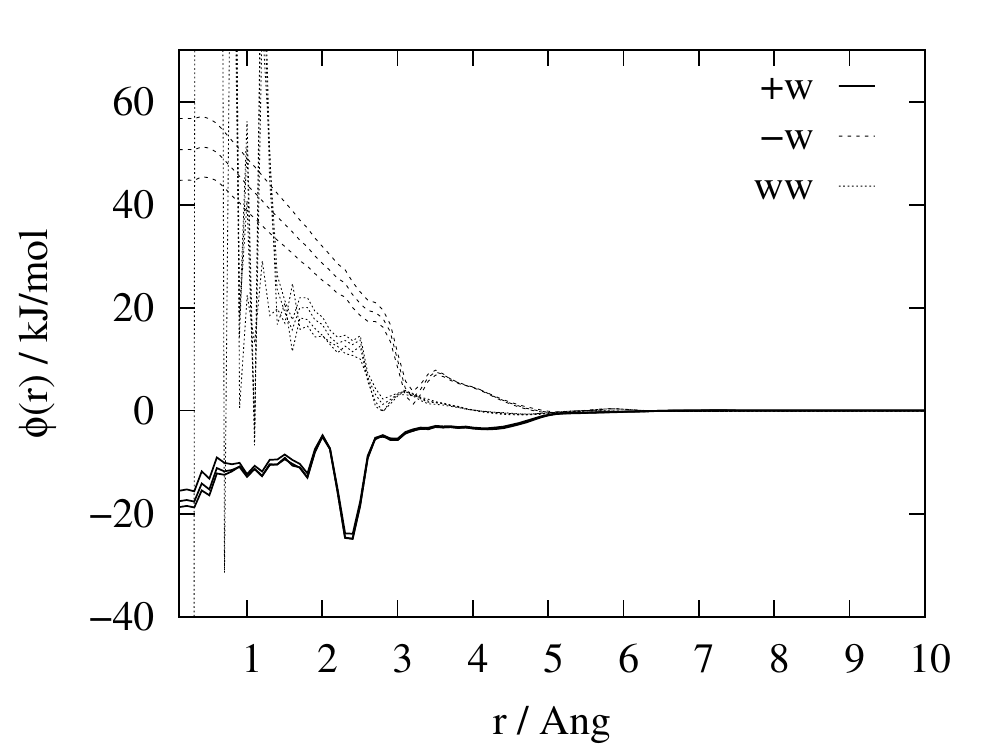}
\caption{Direct correlation functions for cation-water, anion-water,
and water-water center to center interactions.
Roughness of the water-water interaction is not physically meaningful, since
it is within the water's excluded volume.}\label{f:direct}  }
\end{figure}

\begin{figure}
{\centering
\includegraphics[width=0.45\textwidth]{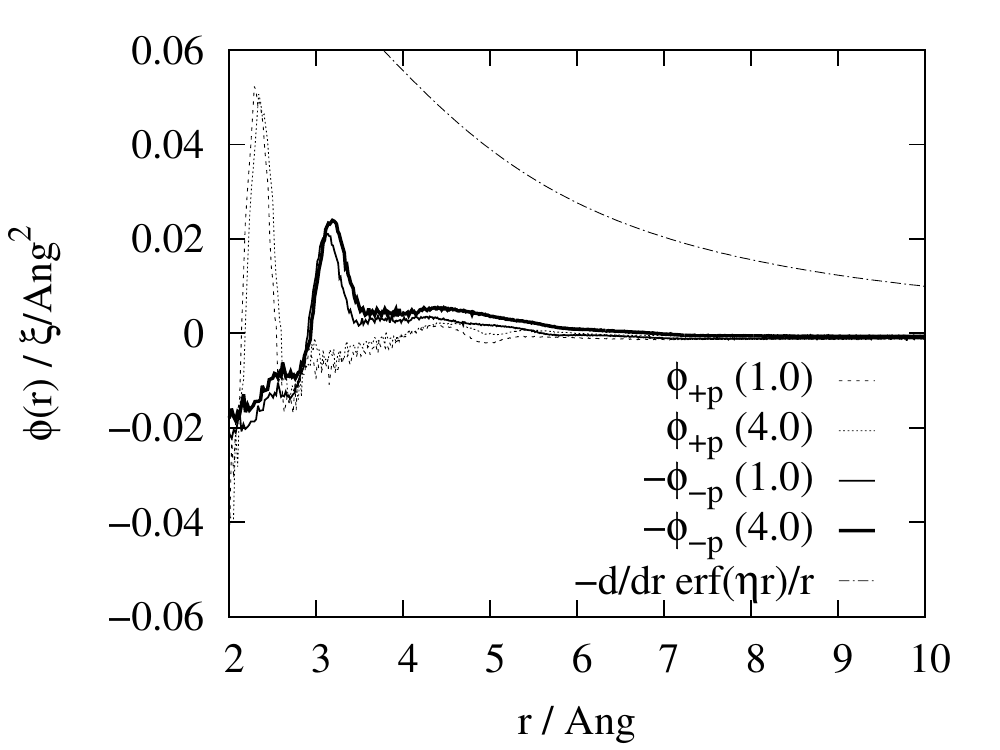}
\caption{Ion-dipole direct correlation functions for 1M and 4M salt concentrations,
scaled by the Coulomb constant, $\xi = 1/4\pi\epsilon_0$.
The long-range Coulomb expression given by the dashed-dotted line
at the bottom was subtracted from each of the lines using
$\eta_{+p} = 1/1.9$\AA{} and $\eta_{-p} = 1/2.3$\AA{}.  The only
major feature that remains is a peak around the distance of
the first maximum in $g(r)$ (Fig.~\ref{f:ions}).}\label{f:ionwat} }
\end{figure}

\begin{figure}
{\centering
\includegraphics[width=0.5\textwidth]{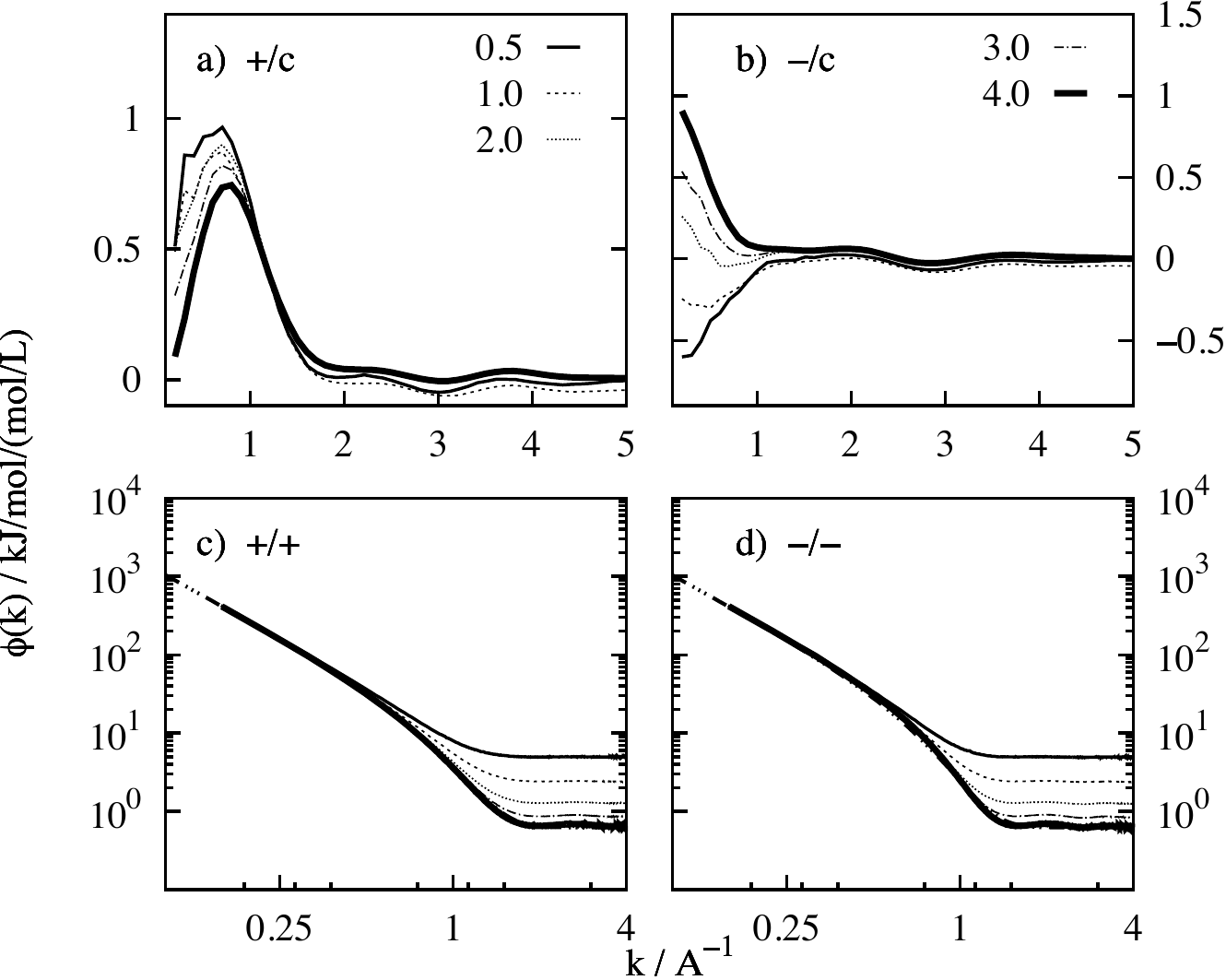}
\caption{Electrolyte direct correlation functions in reciprocal space
for varying salt concentration.
Panels a and b show the average interaction of anions and cations with
all other ions: $(Q^{-1}_{++} + Q^{-1}_{+-})/2 - 1/2\beta \rho$ and
$(Q^{-1}_{--} + Q^{-1}_{+-})/2  - 1/(2\beta4\text{M})$, respectively.
Panels c and d show $Q^{-1}_{++}$ and $Q^{-1}_{--}$ (resp.) on a log-log scale near the origin.
The light gray line is $e^{-k^2/4\eta^2}/\epsilon_0 k^2 + 1/2\beta \rho$ (cf. Eq.~\ref{e:iQLR})
and fits to $1/\eta_{++} = 2.2$\AA{} and $1/\eta_{--} = 2.65$\AA{}.}\label{f:recip} }
\end{figure}

\begin{figure}
{\centering
\includegraphics[width=0.49\textwidth]{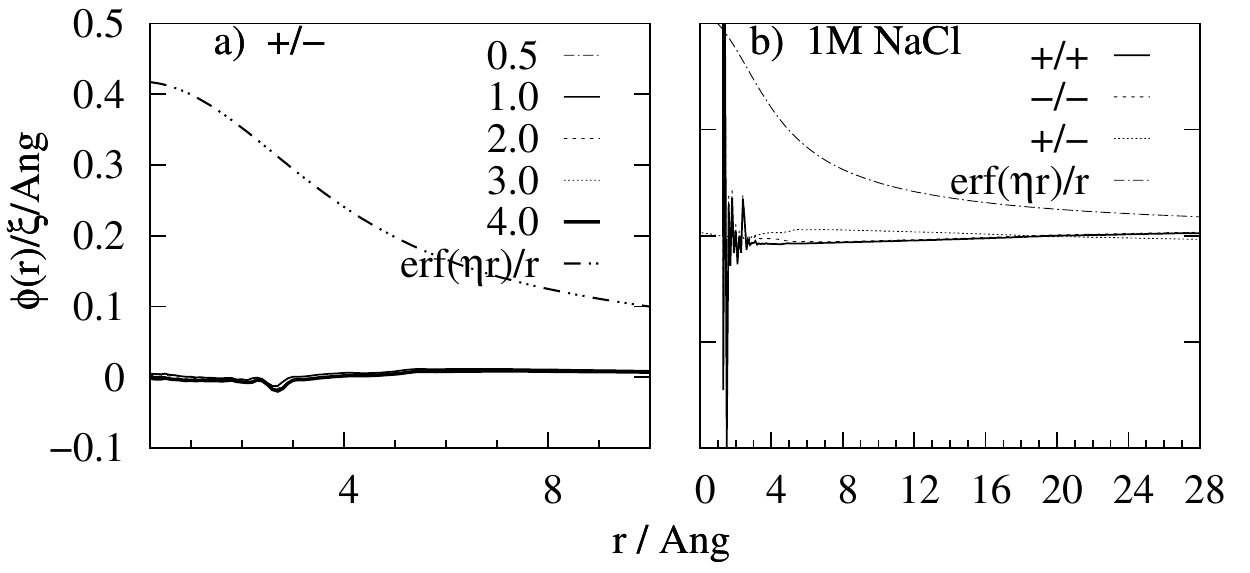}
\caption{Ion center of mass direct correlation functions scaled by the Coulomb constant,
$\xi = 1/4\pi\epsilon_0$.  Panel a shows concentration dependence of cation-anion interactions
(using $\eta = 1/2.7$\AA{}), while panel b compares direct correlation
functions between like and unlike ions.
In every case, the curves are drawn after subtracting the screened Coulomb
expression appropriate for each pair (dashed-dotted lines).
The screening distances used for like ions were the same as in Fig.~\ref{f:recip}.}\label{f:ions2} }
\end{figure}

\begin{figure}
{\centering
\includegraphics[width=0.49\textwidth]{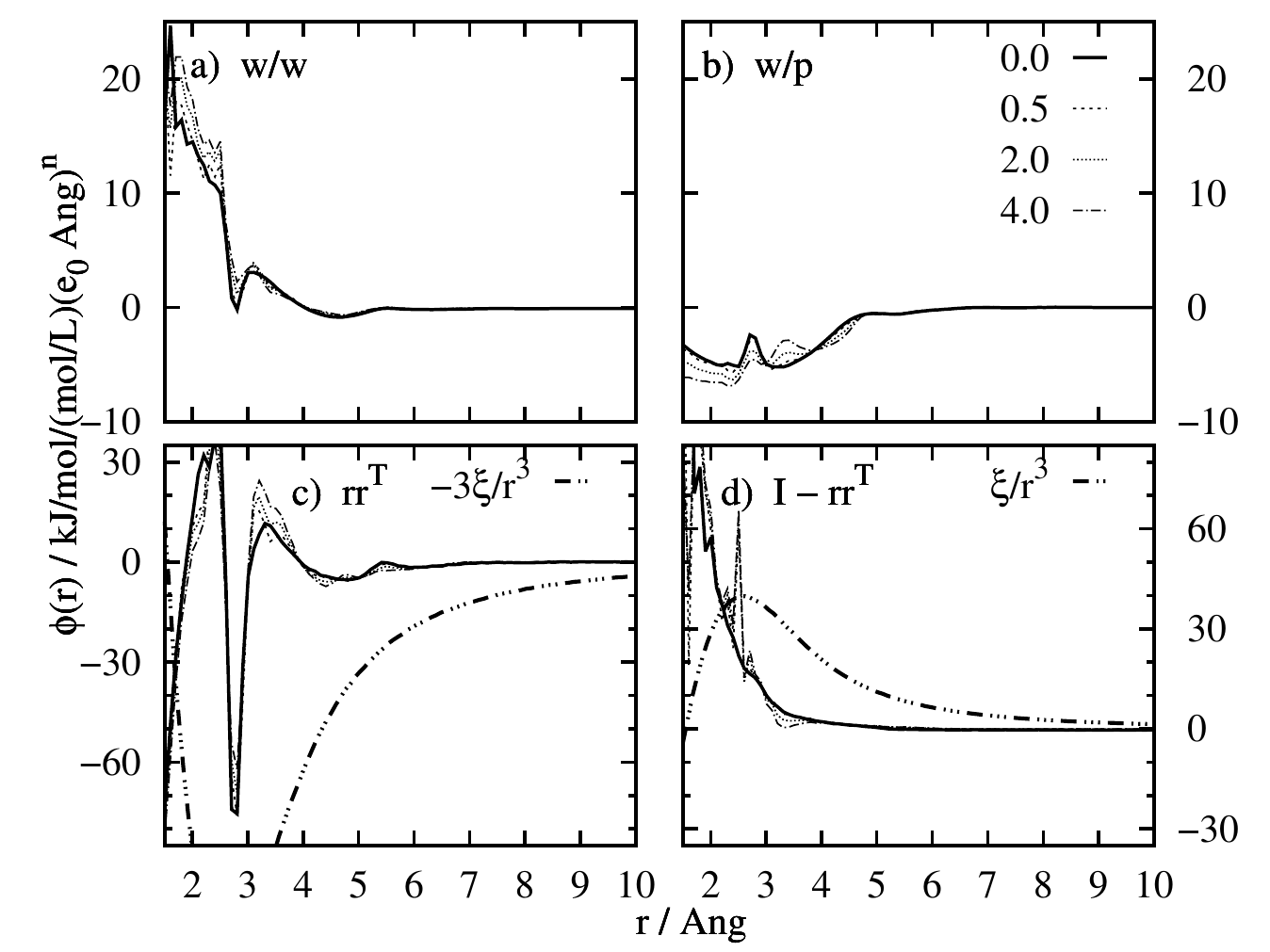}
\caption{Water-water direct correlation functions.
Panel a shows center to center interactions, b shows center to dipole
interactions (positive if the water dipole faces the separation distance),
and panels c and d show co-linear and perpendicular components of the dipole-dipole interaction.
The last two curves are drawn after subtracting the appropriate long-range Coulomb
expression (indicated by the labeled dash-dotted line).}\label{f:wat2} }
\end{figure}

  Figures~\ref{f:ions}-\ref{f:wat2} show the ion-ion, ion-water, and water-water
radial distribution functions and their transformations as a function of salt concentration.
The direct interaction functions in real-space (Figs.~\ref{f:direct}, \ref{f:ionwat}, \ref{f:ions2}, and~\ref{f:wat2})
follow screened Coulomb electrostatic laws so closely that we have subtracted out the screened forms
and plotted the differences instead.
Subtracted parts are shown so that the two curves can be visually added to get
the total $Q^{-1}$.
Ion-water (Figs.~\ref{f:direct} and~\ref{f:ionwat}) and ion-ion interactions (Fig.~\ref{f:ions2}) show negligible concentration dependence.

  Figure~\ref{f:recip} shows the ion-ion interactions in reciprocal
space to have the expected, ideal, $1/\rho$
contribution and the $k^{-2}$ divergence near the origin.\cite{patta93}
The $k \to 0$ limit in Fig.~\ref{f:recip}a and~\ref{f:recip}b can be used to find
the Kirkwood-Buff coefficients for single-ions.\cite{sweer03,sschn13,droge17e}
A log-log plot (Fig.~\ref{f:recip}c,d) shows the $k^{-2}$
divergence transitions to an exponential-like screening around $k=1$\AA{}$^{-1}$ before
flattening out due to the ideal ($1/\rho$) contribution.  These profiles were used to identify
appropriate screening lengths for each ion, resulting in
$1/\eta_{++} = 2.2$\AA{} and $1/\eta_{--} = 2.65$\AA{}. 
Note that Born radii are expected to be temperature and pressure-dependent.
However, our results indicate that they are not sensitive to
salt concentration in the 0-4M range.

  The ion center of mass direct correlation functions in Fig.~\ref{f:ions2}
are surprisingly featureless apart from the $1/r$ interaction.  An unexplainable, small linear trend
shows up in Fig.~\ref{f:ions2}b at long range with a positive slope for like-charged
ion interaction and a negative slope for cation-anion interactions.
At short-range, the like-charged interaction shows numerical noise before a steep,
repulsive increase.  This does not appear in the cation-anion interaction.
This short-range behavior is not directly linked to interaction energies.
In the mean-spherical approximation, it is treated as a fitting parameter to
achieve excluded volume in $g(r)$.

  Fig.~\ref{f:wat2} returns our focus to the number and dipole
density response behavior of water.  The dipole-dipole interaction energies
for co-linear aligned and in-plane parallel orientations (Fig.~\ref{f:wat2}c,d, respectively)
both follow the expected trends.  Within this dipolar response function description,
local effects only appear within the first 6\AA{}.  Interesting salt-concentration
effects show up in the the shape of the water solvation
shell around a water fixed 
at 
the origin (Fig.~\ref{f:wat2}b).  This may be linked to electrostriction
and specific-ion effects.\cite{vmazz17}

\subsection{ Approximate long-range limiting laws}\label{s:limit}

  We show the utility of the theory above by deriving simple
analytical expressions for long-range
ionic chemical potentials (Eq.~\ref{e:mu2}) using a ``screened MSA''
(for a 1:1 electrolyte with equal cation and anion concentrations, $\rho$),
\begin{equation}
Q^{-1}(k) = \begin{bmatrix}
\rho^{-1} + f & -f & i k^T f \\
- f & \rho^{-1} + f & - i k^T f \\
-i k f & i k f & \chi
\end{bmatrix}
,\label{e:iQLR}
\end{equation}
where $\chi = I_3 \beta/\rho_w \alpha + f k k^T$ is the lattice result,
containing contributions from water polarizability
and the screened dipole-dipole interaction energy.
To properly express the energy contribution from polarizing the surface,
the dipole-dipole energy at $k=0$, should be replaced by the
depolarization tensor ($\chi(0) = I_3\beta/\rho_w\alpha + \beta J$).\cite{vball14}
Eq.~\ref{e:iQLR} is essentially a mean-spherical approximation, and
is a caricature of the low-$k$ behavior found in Sec.~\ref{s:sim} with
only one $\eta$ value. 

  This ansatz is simple, supported by our simulation results 
and yields both the Born theory and a mean-spherical approximation-like
Debye-H\"{u}ckle theory as limits.
However, it is not intended to be used in practice, since Ref.~\citenum{nrft}
contains analytical expressions for $Q$ that match our simulation results
for water much more closely.

\subsection{ Born Limit}

  The developments above make re-derivation of Eq.~\ref{e:Eeff} a simple
matter of removing charge-ion interactions.  To accomplish this, delete
the top two rows and columns from $Q^{-1}$ in Eq.~\ref{e:iQLR}
and in Eq.~\ref{e:s2}.  The result is,
\begin{align}
\beta^2 \sigma^2_\text{LR,Born} &= z_\alpha^2 \braket{k} { f^2(k) 
 \chi^{-1} } {k} \notag \\
 &= \frac{z_\alpha^2}{V} \sum_k f(k) - \bar f(k)
 ,\label{e:Born}
\end{align}
which introduces a definition for the scaled interaction energy,
\begin{equation}
\bar f(k) \equiv f / (1 + e^{-k^2/4\eta^2} \alpha \rho_w /\epsilon_0) \equiv f / \epsilon_r(k)
.\label{e:barf}
\end{equation}
The last equality also defines the ``dielectric function,'' $\epsilon_r$, of the lattice model.
It reduces to Eq.~\ref{e:epsr} when $\eta \to \infty$.
The $k^{-2}$ divergence at $k=0$ integrates
to a contribution that scales as $1/V$.

  Because the solvent density response is purely along the longitudinal ($\hat k$)
direction, it can be written in two forms,
\begin{equation}
\Delta \psi_p = i z_\alpha f \chi^{-1} k = \frac{i z_\alpha k}{k^2} \left(1 - 1/\epsilon_r \right)
.\label{e:caution}
\end{equation}
  It is extremely important to notice the difference between $1/\epsilon_r(k)$,
defined for mathematical convenience in Eq.~\ref{e:barf},
and the indirect dipole-dipole correlation function, $\chi^{-1}$,
which gives the both transverse and longitudinal contributions to the orientation
response.  Both these correlations are long-ranged, while their inverses appear
short-ranged.  However, $\epsilon_r$ is only short-ranged in real-space when there are
no other interactions besides electrostatic ones (hence the $k^{-1}$ prefactor),
while $\chi(r)$ is not short-ranged until the electrostatic $1/r^3$ term is subtracted.

\subsection{ Debye-H\"{u}ckel Limit}

  Maximizing only the dipoles, $\Delta \psi_p$, in Eq.~\ref{e:R2}
leads to the linear response equation,
\begin{equation}
Q^{-1}_{pp} | \Delta \psi_p \rangle + Q^{-1}_{pq} | \Delta \psi_q \rangle + | \beta \phi_p \rangle = 0
,\label{e:dip}
\end{equation}
with obvious notation for 2- and 3-component sub-blocks of Eq.~\ref{e:iQLR}.
Replacing this back in Eq.~\ref{e:R2} and simplifying
leads to a charge-only system with $Q^{-1}_{qq}$
scaled by the effective dielectric (Eq.~\ref{e:barf}),
\begin{align}
{\bar Q}^{-1}_{qq}(k) &= 
\begin{bmatrix}
\rho^{-1} + \bar f & -\bar f \\
- \bar f & \rho^{-1} + \bar f \\
\end{bmatrix}
,\label{e:iQDH} \\
\text{and  }
\beta \bar \phi_q &= [\bar f, -\bar f]^T .
\end{align}

  Inverting Eq.~\ref{e:iQDH} yields the ionic charge density
induced by an external potential from a charge at the origin,
\begin{equation}
\Delta \psi_q(k) = - z_\alpha \left(
  \frac{\rho \bar f}{1 + 2\rho \bar f} - \delta_k \right)
\begin{bmatrix} 1 \\ -1 \end{bmatrix}
.\label{e:resp}
\end{equation}
Ref.~\citenum{aniko12} also found Eq.~\ref{e:resp} by applying the
random phase approximation to a fluid of screened, soft-core ions.
They obtain a closed form for its Fourier transform ($h(r)$)
and show that it predicts a cross-over from exponential to damped oscillatory
behavior at high ionic strength.

  Collecting all the terms in Eq.~\ref{e:R2} created by our procedure,
the fluctuations in the solvation potential
of an ion with charge $z_\alpha$ is given by the sum,
\begin{align}
\sigma^2_\text{LR} &= \sigma^2_\text{LR,Born} + \sigma^2_\text{LR,DH} \\
\beta^2 \sigma^2_\text{LR,DH} &= \frac{z_\alpha^2}{V} \sum_{k\ne 0} \frac{2 \rho \bar f^2(k)}{1 + 2\rho \bar f}
 .\label{e:sDH}
\end{align}
Using the relations for the inner product developed in Appendix~\ref{s:ip},
the summation in Eq.~\ref{e:sDH} goes over to the inverse Fourier transform
as the cell volume increases,
\begin{align}
\beta^2 \sigma^2_\text{LR,DH} &\to \frac{\beta z_\alpha^2}{(2\pi)^3 \epsilon_r\epsilon_0}
\int_{0^+}^{\infty} \frac{4\pi \kappa^2 e^{-k^2/2\eta^2}}{k^2 + \kappa^2 e^{-k^2/4\eta^2}} \; dk \notag \\
&= \frac{\beta z_\alpha^2 \kappa}{4 \pi^2 \epsilon_r\epsilon_0}
\int_{-\infty}^{\infty} \frac{e^{-k^2 u^2}}{k^2 e^{k^2 u^2} + 1} \; dk
.\label{e:sDH2}
\end{align}
Here $\kappa^2 \equiv 2\beta\rho/\epsilon_r \epsilon_0$ and $u^2 \equiv \kappa^2/4\eta^2 = \kappa^2 R_\text{B}^2/\pi$.  When $u$ goes to zero (zero ion radius),
the integral becomes $\pi$, and we recover the classical Debye-H\"{u}ckel result for the
solvation free energy using Eq.~\ref{e:sDH2} with Eq.~\ref{e:mu2}.

  For finite $\eta$, $\epsilon_r$ is $k$-dependent, and
the integral is strictly smaller than $\pi$.  It decreases with increasing
$\kappa R$ -- making ionic solvation less favorable.
We calculated the integral numerically and verified that it compares well
with the mean spherical approximation result for the electrostatic free energy component with
ion radius $R_B$.\cite{lblum75}

\section{ Discussion}\label{s:disc}

  The assumption in Eq.~\ref{e:iQLR} used to derive Born and modified DH laws is essentially
the random phase approximation (RPA) of fluid density functional theory.
A similar derivation has recently been presented (starting from the RPA)
by Frydel and Ma.\cite{dfryd16}  They do not include screening and thus arrive at the linearized
Poisson-Boltzmann equation.  Earlier work showed the utility of this approximation
for describing charge reversal near a charged surface.\cite{dfryd13}
In later work, they caution that the RPA, being equivalent
to a variational Gaussian approximation, has the same weaknesses as a mean-field approximation
and loses applicability at strong coupling.\cite{yxian17}
The primary difference in this work is that we do not use the adiabatic connection to
charge up the whole system at once, but only calculate the effect of
adding a single ion in a self-consistent way.  This frees approximations like
Eq.~\ref{e:R2} from having to represent the density of the entire system.
The RPA works well because the solute potential, $\Phi_\alpha$,
is concentrated at long range, near $k=0$.

  We view the focus on addition of a single molecule to a pre-existing fluid as the
major reason for success of the present theory.  An entire fluid constructed from
soft ions has very different properties.  It has been investigated as the
``ultrasoft restricted primitive model'' (URPM) of penetrable electrolytes
such as charged polymers.\cite{aniko12}
The screening prevents Coulomb collapse in the URPM, but allowing ion
overlap removes the competition between long-range attraction and short-range excluded
volume underpinning most density functional closures,
explaining their difficulty in predicting its phase diagram.

  It is hard to compare our simulation results with the body of literature
on the Hubbard-Stratonovich transformation, which is a Fourier instead of a Legendre
transform.
The major difficulty is technical.  Since the transformation introduces an imaginary,
auxiliary field with an infinite number of degrees of freedom, most of its expressions
are related to distributions of the imaginary potential and cannot be easily
compared to molecular simulations.  Analytically, the theory is commonly used in combination
with a mean-field type expression for the potential or density functional
which can result in various modified Poisson-Boltzmann theories depending
on this approximation.\cite{sbuyu13,alevy13,jpujo15}
When employing the variational Gaussian approximation within the statistical
field theory, the results of Sec.~\ref{s:limit} appear almost exactly
with the same conclusions.\cite{zwang10,jmart16}  Nevertheless, the picture there is of the URPM model
and is more difficult to extend to non-pairwise interactions.
In addition, entropic terms similar to the
primitive available volume approximation ($-\ln(1 - V_0 \rho)$)
have been successfully used to account for the free energy of cavity formation
in related works.\cite{iboru97,pkoeh09}


  The physical picture of solvent response gained from Fig.~\ref{f:wat2}
is that the ``local electric field,'' which determines the force felt by
a solvent dipole has special contributions from the first two or three solvation layers.
Beyond that, it has the expected form of an integral over dipole-dipole electrostatic
$1/r^3$ terms.  Two maxima surround a sharp minimum in $\chi^0$
near the first peak of the water-water radial distribution (Fig.~\ref{f:wat2}c)
-- so that the dipole-dipole interaction is slightly more repulsive for
waters deviating from the optimal distance.
This behavior likely reflects quadrupolar forces causing correlations between the position and
orientation angle of first-shell waters.  The isotropic part of the local electric field
(that scales with the cosine of the water-water angle, $\mu\cdot\mu$) in
Fig.~\ref{f:wat2}d mainly follows a typical screened form,
but shows an extra tendency for
waters closer than 3\AA{} to take antiparallel orientations.\cite{dmart08}

  For ion-water association, Fig.~\ref{f:ionwat} shows a single peak where
the usual ion-dipole interaction is strengthened, while Fig.~\ref{f:direct}
shows a local minimum for the center-to-center interaction.  This strengthening occurs
for both ions and both interactions right at the maximum of the ion-water
radial distribution function.  The extra dipolar response is direct evidence
for the hypothesis that the first solvent dipole layer at the solute/solvent interface
should be scaled from the Maxwell theory.\cite{mheyd12}  The extra
center-to-center interaction provides an energetic basis for electrostriction.
Both are associated with specific-ion effects.\cite{wwach05,tbeck13}

\section{ Conclusion}

  The limiting cases studied throughout Sec.~\ref{s:limit} establish the utility
of this theory, and serve as a guide for translating the language of
dielectric polarization into structural, energetic terms.
Future work should explore the key features of the correlation functions
responsible for experimental evidence of specific ion effects.

  We have taken one step, showing the
direct correlation function found from simulations has a surprisingly good agreement
with the random phase and mean spherical approximations.
They are indistinguishable past the second solvation
shell of water.  Even within the first and second solvation shells, the effective energy only oscillates
about the error-function screened form.  One way to rationalize the far-field result is to note
that long-range electrostatic ordering (and charge density fluctuations) does not depend sensitively
on local charge ordering.  Rigorously, higher-order multipoles have a rapidly decreasing interaction
radius.

  We reiterate here the importance of distinguishing between the ideal dielectric continuum
theory, for which $\epsilon_r$ can be usefully defined, and real solutions containing additional
local interactions besides electrostatics.  As Eq.~\ref{e:compar} and~\ref{e:caution} illustrate,
the inverse correlation function is more informative than an inferred dielectric function.

  This observation helps give a simple mathematical structure to a nonlocal response theory
for solution density response to the solute's molecular field.  It is well-known that a majority
of the solvation free energy in polar liquids is due to long-range solvent electrostatic response.
Most dielectric theories spend a lot of effort finding a solute-solvent
boundary that makes bulk dielectric polarization theories give accurate free energies.
The density response paradigm parallels density functional theories by focusing
on reproducing average solvent dipole (and ionic) densities conditional on interaction with the solute.
Eq.~\ref{e:muLR} then provides the excess chemical potential exactly.
Using a screened electrostatic interaction removes non-physical issues
with singularities of the Coulomb potential at the origin, while capturing long-range
contributions.

  The same division into short and perturbative long-range interactions is a crucial step in
local molecular field (LMF) theory,\cite{ychen06,jrodg08,rrems16}.
While LMF theory takes an additional step to find a long-range
Coulomb field self-consistently,
the present work has the more modest
goal of describing only single-solute energetics.
Our connection with solution spatial response functions allows for detailed tests of the theory,
and suggests simple generalizations to self-consistent
spatial and time-dependent problems, for which response data
are already available.\cite{bbagc89,jsala01}

  The technical foundation for the theory applied the Gibbs conditioning principle
to transform the grand potential into an exact expression for the solvation free energy.\cite{cleon02}
This appears to be a novel route for eliminating
the Hubbard-Stratonovich transform that proceeds directly to a computable density functional,
and for expressing the free energy in conventional density functional theories.\cite{akova00}
Although few density functionals treat molecular orientation in an angular expansion,\cite{droge15}
many ideas from both integral equations and density functional theory
can still be carried over into the present formalism.
Another obvious addition is to include the contribution of energetic degeneracy
of solvent dipole orientations belonging to a given dipole density in Eq.~\ref{e:R2}.
This would begin to address the dispersion contribution identified
for the lattice model in Eq.~\ref{e:disp}.

  There are many important questions which might be addressed with this theory.
We have not attempted to calculate the second step in Fig.~\ref{f:cyc},
which involves forming a cavity at the center of a screened field.
It has, however, been explicitly computed by others\cite{rrems16}
who have shown that cavity formation in water creates a positive relative
potential at the center.\cite{hashb06,droge10,mtabr17,tduig17}
That asymmetry lowers the free energy of anionic solvation and,
by path equivalence, makes cavity formation in step 2 of Fig.~\ref{f:cyc} easier
after polarizing water with a negative potential.
Including both polarization and cavity formation in the first step
(by adding an occupancy constraint)\cite{ghumm96}
provides a possible route to combining and comparing these ideas.
The limit, $Q^{-1}(0)$, provides $\partial \beta\mu/\partial \rho$ via
the Kirkwood-Buff  theory,\cite{droge17e} and, given a predictive theory for $Q$,
gives a more traditional route to solvation free energies.\cite{rperr88,sweer03,sschn13}
Further, cluster expansion of Eq.~\ref{e:Z} could be used to include explicit solvent molecules
near the solute as a means to systematically improve approximations to Eq.~\ref{e:muLR}
in a quasi-chemical way.\cite{pdt6,lprat03,droge12}

\section*{Acknowledgments}
This work was supported by the USF Research Foundation.

%

\appendix
\section{ Inner Product Definition}\label{s:ip}

  The definition of the inner product in Eq.~\ref{e:uLR} is straightforward, but extremely
useful both for simplifying our notation and performing computations.
Denote a vector of $\nu$ functions each in $\mathbb R^3 \to \mathbb C^d$, as $f$
and another vector as $g$.  The inner product is the sum-integral,
\begin{equation}
\inner{f}{g} = \sum_{\alpha = 1}^\nu \int_V f_\alpha(r)^\dagger g_\alpha(r) \; dr
,
\end{equation}
where the integration ranges over the 3D unit cell with volume, $V$.
The dagger denotes the complex-transpose and is used so that $f_\alpha(r)^\dagger g_\alpha(r)$
is a scalar.  Note that $\inner{f}{g}$ is the complex conjugate of $\inner{g}{f}$,
motivating our adoption of Dirac brackets.

  Defining Fourier transforms of each function as,
\begin{equation}
\mathcal F[f_\alpha](k) \equiv \int_V e^{-ik\cdot r} f_\alpha(r) \; dr,
\end{equation}
the Fourier-Plancherel theorem proves the equivalence of the inner product with the infinite sum,
\begin{equation}
\inner{f}{g} = \frac{1}{V} \sum_{\alpha = 1}^\nu \sum_k \mathcal F[f_\alpha](k)^\dagger \mathcal F[g_\alpha](k)
.\label{e:iFT}
\end{equation}
In Eq.~\ref{e:iFT}, the summation is over the reciprocal lattice, $k \in \{2\pi L^{-1} z | z \in \mathbb Z^3\}$,
where $L$ is a $3\times 3$ matrix whose rows are unit cell translation vectors.
In the text we use the same inner product symbol for both representations, so
$\inner{\mathcal F[f]}{\mathcal F[g]} \equiv \inner{f}{g}$.

  In the limit of infinite volume, the reciprocal lattice sum goes over to,
\begin{equation}
\lim_{V\to\infty} \frac{1}{V} \sum_k r(k) = \int_{\mathbb R^3} \frac{r(k) \; dk}{(2\pi)^3}
.
\end{equation}

  This work uses hats (as in $\Delta \hat U_\alpha$)
to denote functions which depend on the system microstate.
This is distinct from their usual quantum-mechanical interpretation
as operators.  Mathematically, the hat just denotes a random variable here.

  Operator notation is defined similarly to the inner-product notation,
\begin{align}
\braket{f}{M}{g} &= \sum_{\alpha\gamma} \int_V \int_V f_\alpha(r) M_{\alpha\gamma}(r - r') g_\gamma(r') dr dr' \\
&= \frac{1}{V} \sum_{\alpha\gamma} \sum_k \mathcal F[f_\alpha](k)^\dagger
\mathcal F[M_{\alpha\gamma}](k) \mathcal F[g_\gamma](k)
.
\end{align}

\end{document}